# Uncovering Hidden Variables: A Physics Classroom Activity on Correlation and Causation

Álvaro Suárez[1], Marcelo Vachetta[2]

[1,2]Consejo de Formación en Educación, Instituto de Profesores Artigas, Montevideo, Uruguay

[1]alsua@outlook.com, [2]mvachetta@gmail.com

## Abstract

A classroom activity for high school physics students is presented to explore the distinction between correlation and causation. Using data linking ice cream sales to drowning deaths, presented within a fictional news article, students analyze the relationship, create plots, and identify temperature as the hidden variable explaining the correlation. The activity fosters critical thinking by engaging students in scientific reasoning, illustrating how data representation influences conclusions, and supporting the development of scientific literacy and data interpretation skills.

## 1. Introduction

One of the main goals of physics education is to develop in students the ability to think critically about data and scientific models [1]. This skill is key in everyday life, as it allows one to interpret data accurately, evaluate whether conclusions are based on evidence, and distinguish between meaningful patterns and random occurrences [2].

Although "correlation does not imply causation" it is common to make mistakes when interpreting relationships between variables, either due to bias or the human tendency to seek causal explanations. The history of science is filled with examples where causality was incorrectly inferred from correlation. In today's society, with the rise of fake news, misinterpretation, and data manipulation, educating students about scientific reasoning is more important than ever. In this paper, we present an activity aimed at high school physics students as part of an introductory module on scientific work. The learning objectives of this activity are to adequately plot a set of data, to extract relevant information from these data, and to clearly distinguish between correlation and causation. A brief overview of some types of correlation that do not imply causation follows, followed by a description of the activity, the results obtained, and finally the conclusions.

## 2. Correlation and causality

In science and in everyday life, it is common to encounter situations where two correlated variables are mistakenly interpreted as having a cause-and-effect relationship. The simplest case of this error is correlations between quantities that have no logical connection, known as "spurious correlations", which are strong examples that "correlation does not imply causation". Hundreds of such examples can be found on the Internet, one of the most popular being the correlation between the number of people who drowned in a swimming pool in the United States and the number of movies in which Nicolas Cage has starred [3].

Another type of correlation, often misinterpreted as implying causation, occurs when two variables change simultaneously, and a causal relationship is incorrectly inferred. This type of error is common in several areas of physics [4, 5] and can lead to conceptual difficulties, as in the case of Faraday's and Ampere-Maxwell's laws [5]. According to this misinterpretation, Faraday's law implies that a time variation of a magnetic field generates an electric field, while Ampere-Maxwell's law implies that a time variation of an electric field generates a magnetic field. However, all terms in these laws are evaluated at the same instant of time, and therefore they do not imply cause and effect relationships, since none of these fields can be the cause of the other [6, 7, 8].

A third type of correlation, which does not imply causation, occurs when there is a third "hidden" variable that is the common cause of the two correlated quantities. A famous case was published in Nature [9] where the causes of myopia in children under the age of two were studied. It was found that those who slept with night lights were more likely to develop this visual disorder, leading to the conclusion that sleeping with lights on could cause myopia. However, publications soon appeared that refuted these conclusions [10]. Although the correlation was apparent, it was not a causal relationship. The hidden variable in this case was the parents' myopia—parents with poor vision were more likely to leave the lights on at night. Therefore, the most likely cause of myopia in children was genetic inheritance.

## 3. The activity

For our activity, we tasked the students -organized into five groups of four or five members- with analyzing a potential correlation between two quantities influenced by a hidden variable. Specifically, they explored the relationship between the number of drownings in swimming pools and ice cream sales. The task was framed as a report from a fictitious news article that claimed to have discovered a connection between these two variables, including monthly data on swimming pool drownings for the year 2022 in US. To create a more engaging context, the activity included images of newspaper front pages generated with AI tools. To obtain the data, we asked Microsoft's AI chat to provide information on swimming pool drownings in the United States for 2022 (Table 1). While we did not cross-check the accuracy of these data with other sources, our primary goal was to ensure the dataset exhibited a clear seasonal trend, such as a higher number of drownings in summer months. The activity did not aim to verify the authenticity of the data, as the news report itself was explicitly fictional.
Each group of students was given a whiteboard to complete the activity [11]. The activity as presented to the students is transcribed below.

*"DEADLY ICE CREAM"*

In the fall of 2023, disturbing data became known about a possible link between ice cream consumption and the number of drownings in US swimming pools during 2022. International newspapers echoed this news. The data that generated concern among researchers, swimmers and consumers of the sweet elixir were presented in the following table.

a) Based on the data collected, a relationship between the two variables may be suspected. Draw a plot that allows you to investigate this relationship.
b) Can you draw any conclusions from the plot?
c) Some sensationalist newspapers published "the danger of combining ice cream and swimming pools". If you owned an ice-cream factory, what arguments would you use to defend your company?

At the end of the activity, students completed the following exit ticket: Write a paragraph about what you learned in today's lesson.

| Month | Ice Cream Sales (millions of dollars) | Swimming Pool Drowning Deaths |
|---|---|---|
| 1 | 200 | 20 |
| 2 | 190 | 17 |
| 3 | 210 | 19 |
| 4 | 240 | 25 |
| 5 | 270 | 35 |
| 6 | 450 | 75 |
| 7 | 480 | 90 |
| 8 | 460 | 85 |
| 9 | 320 | 50 |
| 10 | 280 | 40 |
| 11 | 220 | 22 |
| 12 | 210 | 20 |

Table 1. Data table obtained with AI.

## 4. Implementation results

When the students received the assignment, they were initially skeptical, asking multiple times whether both the data and the news report were real. As they began working, they had to determine which quantities to plot in order to investigate the relationship between the variables. Three of the teams chose to plot the number of drownings versus ice cream sales (Figure 1), while the other two teams plotted both drownings and ice cream sales on the same set of axes, with both variables versus the month of the year (Figure 2). All the groups quickly recognized the existence of a correlation, which surprised them. However, they did not all agree on whether or not there was a causal relationship and formulated different hypotheses.

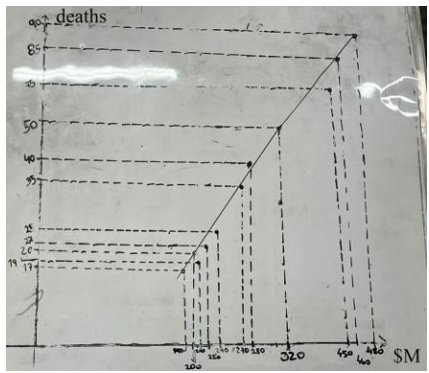

Fig. 1. Plot of the number of drownings versus ice cream sales.

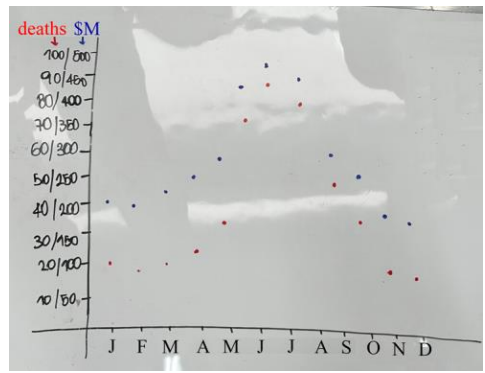

Fig. 2. Plot of the number of drownings and ice cream sales versus the month of the year.

The groups that plotted the number of deaths and ice cream sales over the months were able to see that although the two variables were correlated, the underlying cause was the variation in temperature over the year. One of the teams concluded:

*"In conclusion, the months associated with more deaths are summer. Our hypothesis is that ice cream is not related to deaths, but drowning deaths are related to the increase in heat due to summer".*

On the other hand, the remaining teams failed to identify the "hidden variable", probably because they plotted the number of deaths as a function of ice cream sales. This led them to formulate different hypotheses to explain the relationship between the two variables. One of these teams concluded:

*"While ice cream is associated with deaths, it is also a dessert that helps to cool down on hot days. Perhaps one solution is to wait a certain amount of time before entering the pool after eating ice cream."*

At the end of the allotted time, the groups presented their findings during a plenary session using the blackboard. Each group had developed different hypotheses regarding the possible cause of the correlation between drownings and ice cream sales, and they presented the data in various ways. This led to a rich discussion, allowing for a deeper understanding of the interpretation of the data and the distinction between correlation and causation.

At the end of the activity, each student individually completed an exit ticket. Analysis of the responses confirmed that the activity was productive and that students were able to understand that correlation does not necessarily imply causation. Let's look at two student responses:

*E1: "I learned that two things can be related, but one does not necessarily cause the other".*

*E2: "I learned that you have to analyze all the data well before drawing a conclusion and not get confused by 'coincidences' like the deaths and the ice cream activity. The number of deaths was related to the sale of ice cream, but in reality it was a coincidence and the most logical explanation is that there are more drowning deaths in the summer when there are more ice cream sales".*

As we can see, students emphasize that not all correlations imply causation and stress the importance of careful analysis of data.

Although it was not part of our original objective, given the results obtained, we consider this activity a valuable opportunity for educators to further explore the inherent subjectivity involved in interpreting data. The two types of graphs created by the students led to different, yet logical, conclusions—both of which were valid representations of the data. This demonstrates that drawing "evidence-based conclusions" is not always purely objective or "true". Rather, conclusions depend on how data are represented and analyzed. Encouraging students to consider multiple perspectives on a problem can foster a deeper understanding and more robust conclusions, helping them recognize how implicit biases and methodological choices shape scientific claims. Furthermore, this reflection provides an opportunity to discuss how such subjectivity, while often unavoidable, can also open the door to data manipulation or misinterpretation.

## 5. Conclusions

Researching, evaluating, and using scientific information to make decisions and take action is a core scientific competency [12] and is becoming increasingly important in an increasingly complex world where we are constantly bombarded with information. In this context, it is crucial that our students are able to distinguish between scientific claims that are evidence-based and those that are not.

Based on the students' responses when implementing the activity presented in this paper, we believe that this can be a valuable resource for promoting scientific literacy in our students by placing them in a situation where they must distinguish and understand the difference between correlation and causation. By engaging students in this type of activity, we encourage the development of critical thinking and contribute to the development of a quality scientific culture that prepares students to successfully face the challenges of analyzing and interpreting information in an increasingly complex world.